\documentclass[12pt]{article}
\usepackage{epsf}
\usepackage{graphicx}

\begin{document}
\title
{Lorentz symmetry violation due to interactions of photons with
the graviton background}
\author
{Michael A. Ivanov \\
Physics Dept.,\\
Belarus State University of Informatics and Radioelectronics, \\
6 P. Brovka Street,  BY 220027, Minsk, Republic of Belarus.\\
E-mail: michai@mail.by.}

\maketitle

\begin{abstract}
The average time delay of photons due to multiple interactions
with gravitons of the background  is computed in a frame of the
model of low-energy quantum gravity by the author. The two
variants of evaluation of the lifetime of a virtual photon are
considered: 1$)$ on a basis of the uncertainties relation (it is a
common place in physics of particles) and 2$)$ using a conjecture
about constancy of the proper lifetime of a virtual photon. It is
shown that in the first case Lorentz violation is negligible: the
ratio of the average time delay of photons to their propagation
time is equal approximately to $10^{-28}$; in the second one (with
a new free parameter of the model), the time-lag is proportional
to the difference $\sqrt{E_{01}}-\sqrt{E_{02}}$, where $E_{01},\
E_{02}$ are initial energies of photons, and more energetic
photons should arrive later, also as in the first case. The effect
of graviton pairing is taken into account, too.
\end{abstract}

\section[1]{Introduction }
Lorentz invariance is the cornestone of physics of elementary
particles, and a degree of its possible violation is of a great
interest (see the review \cite{400}). Possible Lorentz violation
is often connected in our minds with quantum gravity effects; and
it is almost commonly accepted that these effects should reveal
themselves at the Plank scales of energies and distances. It is
another story that dealing with the Plank scale of distances
suggests that our knowledge of gravity (general relativity) is
true up to this scale \cite{401}; but it is not a proofed fact. I
would like to cite the recent paper \cite{402} as a typical one in
this direction; the authors speak about days or months of
time-lags for photons of GRB's in some theoretical cases.
\par But in my model of low-energy quantum gravity \cite{500},
gravity reveals asymptotic freedom at very short distances
beginning from $10^{-11} - 10^{-13}$ meter for different particles
\cite{116}, i.e. very-very far from the Plank scale. In this
paper, I have computed the average time delay of photons due to
multiple interactions with gravitons of the background in a frame
of the model \cite{500}. The two variants of evaluation of the
lifetime of a virtual photon are considered: 1$)$ on a basis of
the uncertainties relation (it is a common place in physics of
particles) and 2$)$ using a conjecture about constancy of the
proper lifetime of a virtual photon. It is shown that in the first
case Lorentz violation is negligible; in the second one (with a
new free parameter of the model), the time-lag is proportional to
the difference $\sqrt{E_{01}}-\sqrt{E_{02}}$, where $E_{01},\
E_{02}$ are initial energies of photons, and more energetic
photons should arrive later, also as in the first case. The effect
of graviton pairing is taken into account, too.

\section[2]{Time delay of photons due to interactions with gravitons }
To compute the average time delay of photons in the model
\cite{500}, it is necessary to find a number of collisions with
gravitons of the graviton background on a small way $dr$ and to
evaluate a delay due to one act of interaction. Let us consider at
first the background of single gravitons. Given the expression for
$H$ in the model, we can write for the number of collisions with
gravitons having an energy $\epsilon =\hbar \omega$:
\begin{equation}
dN(\epsilon)= {|dE(\epsilon)| \over \epsilon} = E(r)\cdot {dr
\over c} {1 \over 2\pi}  D f(\omega,T)d \omega,
\end{equation}
where $f(\omega,T)$ is described by the Plank formula. In the
forehead collision, a photon loses the momentum $\epsilon/c$ and
obtains the energy $\epsilon$; it means that for a virtual photon
we will have:
\begin{equation}
{v\over c}={{E- \epsilon}\over {E+ \epsilon}}; \ 1- {v\over c}={
2\epsilon \over {E+ \epsilon}}; \ \ 1- {v^2\over c^2}={ 4\epsilon
E \over (E+ \epsilon)^2}.
\end{equation}
\subsection[2.1]{Evaluation of the lifetime of a virtual photon on a basis of the
uncertainties relation} The uncertainty of energy for a virtual
photon is equal to $\Delta E=2 \epsilon$. If we evaluate the
lifetime using the uncertainties relation: $\Delta E \cdot \Delta
\tau \geq \hbar/2 $, we get $\Delta \tau \geq \hbar/4\epsilon$. So
as during the same time $\Delta \tau$ real photons overpass the
way $c \Delta \tau$, and virtual ones overpass only the way $v
\Delta \tau$, we have: $$c \Delta t = c \Delta \tau - v \Delta
\tau,$$ where $\Delta t$ is the time delay, and the last one will
be equal to:
\begin{equation}
\Delta t(\epsilon)=\Delta \tau (1- {v\over c})\geq \hbar/2 \cdot
{1 \over {E+ \epsilon} }.
\end{equation}
The full time delay due to gravitons with an energy $\epsilon$ is:
$dt(\epsilon)=\Delta t(\epsilon) dN(\epsilon)$. Taking into
account all frequencies, we find the full time delay on the way
$dr$:
\begin{equation}
dt\geq  \int_{0}^{\infty} {\hbar \over 2}{E  \over
{E+\epsilon}}\cdot {dr \over c} {1 \over 2\pi}D f(\omega,T)d
\omega.
\end{equation}
The one will be maximal for $E \rightarrow \infty, $ and it is
easy to evaluate it:
\begin{equation}
dt_{\infty} \geq  {\hbar \over 4\pi}{dr \over c}\cdot D \sigma
T^4.
\end{equation}
On the way $r$ the time delay is:
\begin{equation}
t_{\infty}(r) \geq  {\hbar \over 4\pi}{r \over c}\cdot D \sigma
T^4.
\end{equation}
In this model: $r(z)=c/H\cdot \ln(1+z)$; let us introduce a
constant $\rho \equiv {\hbar / 4\pi}\cdot D \sigma T^4/H = 37.2
\cdot 10^{-12} s$, then
\begin{equation}
t_{\infty}(z) \geq \rho \ln(1+z).
\end{equation}
We see that for $z\simeq 2 $ the maximal time delay is equal to
$\sim 40 \ ps$, i.e. the one is negligible.
\par In the rest frame of a virtual photon, a single parameter,
which may be juxtaposed with an energy uncertainty, is $mc^{2}$.
Accepting $\Delta E = mc^{2}$ in this frame, we'll get:
\begin{equation}
t(z) \geq \rho/2 \cdot \ln(1+z)
\end{equation}
with the same $\rho$; now this estimate doesn't depend on $E$.
\subsection[2.2]{The case of constancy of the proper lifetime of
a virtual photon} Taking into account that for a virtual photon
after a collision $ (E^{'}/c)^{2}-p^{'2}> 0,$ we may consider
another possibility of lifetime estimation, for example, $\Delta
\tau_{0}=const$, where $\Delta \tau_{0}$ is the proper lifetime of
a virtual photon (it should be considered as a new parameter of
the model). Now it is necessary to transit to the reference frame
of observer:
\begin{equation}
\Delta \tau =\Delta \tau_{0}/(1- {v^2\over c^2})^{1/2}=\Delta
\tau_{0}\cdot{{E+\epsilon} \over 2\sqrt{\epsilon E}},
\end{equation}
accordingly:
\begin{equation}
\Delta t(\epsilon)=\Delta \tau (1- {v\over c})=\Delta
\tau_{0}\cdot\sqrt{\epsilon/ E}.
\end{equation}
Then the full time delay due to gravitons with an energy
$\epsilon$ is:
\begin{equation}
dt(\epsilon)=\Delta t(\epsilon) dN(\epsilon)= \Delta
\tau_{0}\cdot\sqrt{\epsilon E}\cdot {dr \over c} {1 \over 2\pi}D
f(\omega,T)d \omega,
\end{equation}
and integrating it, we get:
\begin{equation}
dt= \Delta \tau_{0}\cdot\sqrt{ E(r)}\cdot {dr \over c} {1 \over
2\pi}D \int_{0}^{\infty} \sqrt{\epsilon}f(\omega,T)d \omega.
\end{equation}
The integral in this expression is equal to:
\begin{equation}
\int_{0}^{\infty} \sqrt{\epsilon}f(\omega,T)d \omega \equiv {1
\over {4\pi^{2}c^{2}}}\cdot {(kT)^{9/2} \over {\hbar^{3}}}\cdot
I_{6},
\end{equation}
where a new constant $I_{6}$ is the following integral:
\begin{equation}
I_{6} \equiv \int_{0}^{\infty} {x^{7/2}dx \over {\exp{x}-1}} =
12.2681.
\end{equation}
In this model, the energy of a photon decreases as \cite{500}:
$E(r)=E_{0}\exp(-Hr/c). $ The full delay on the way $r$ now is:
\begin{equation}
t(r)= \Delta \tau_{0}\cdot {D \over {8\pi^{3}c^{2}}}\cdot
{(kT)^{9/2} \over {\hbar^{3}}}\cdot I_{6}\int_{0}^{r} \sqrt{
E(r{'})}\cdot {dr{'} \over c}=
\end{equation}
$$ =\Delta \tau_{0}\cdot {D \over
{8\pi^{3}c^{2}}}\cdot {(kT)^{9/2} \over {\hbar^{3}}}\cdot
I_{6}\cdot {2 \over H}\cdot (\sqrt{ E_{0}}-\sqrt{ E(r)}).$$ Let us
introduce a new constant $\epsilon_{0}$ for which:
$${1\over \sqrt \epsilon_{0}} \equiv {D \over {8\pi^{3}c^{2}}}\cdot
{(kT)^{9/2} \over {\hbar^{3}}}\cdot I_{6}\cdot {2 \over H},$$ so
$\epsilon_{0}=2.391 \cdot 10^{-4} \ eV,$ then
\begin{equation}
t(r)= {\Delta \tau_{0} \over \sqrt \epsilon_{0}}\cdot(\sqrt{
E_{0}}-\sqrt{ E(r)})=\Delta \tau_{0}\sqrt{E_{0} \over
\epsilon_{0}}\cdot (1-\exp(-Hr/2c)),
\end{equation}
where $E_{0}$ is an initial photon energy. This delay as a
function of redshift is:
\begin{equation}
t(z) =\Delta \tau_{0} \sqrt{E_{0} \over \epsilon_{0}} \cdot
{{\sqrt{1+z} -1} \over \sqrt{1+z}}.
\end{equation}
\par In this case, the time-lag between photons emitted in one
moment from the same source with different initial energies
$E_{01}$ and $E_{02}$ will be proportional to the difference
$\sqrt{E_{01}}-\sqrt{E_{02}}$, and more energetic photons should
arrive later, also as in the first case. To find $ \Delta
\tau_{0}$, we must compare the computed value of time-lag with
future observations. An analysis of time-resolved emissions from
the gamma-ray burst GRB 081126 \cite{372} showed that the optical
peak occurred $(8.4 \pm 3.9) \ s$ {\it later} than the second
gamma peak; perhaps, it means that this delay is connected with
the mechanism of burst.
\subsection[2.3]{An influence of graviton pairing }
Graviton pairing of existing gravitons of the background is a
necessary stage to ensure the Newtonian attraction in this model
\cite{6}. As it has been shown in the cited paper, the spectrum of
pairs is the Planckian one, too, but with the smaller temperature
$T_{2}\equiv 2^{-3/4} T;$ this spectrum may be written as:
$f(\omega_{2},T_{2})d \omega_{2}$, where
$\omega_{2}\equiv2\omega.$ Then residual single gravitons will
have the new spectrum: $f(\omega,T)d \omega - f(\omega_{2},T_{2})d
\omega_{2}$, and we should also take into account an additional
contribution of pairs into the time delay.
\par We shall have now:
\begin{equation}
dN(\epsilon)= E(r)\cdot {dr \over c} {1 \over 2\pi}  D
(f(\omega,T)d \omega - f(\omega_{2},T_{2})d \omega_{2}),
\end{equation}
and for pairs with energies $2\epsilon:$
\begin{equation}
dN(2\epsilon)= {|dE(2\epsilon)| \over 2\epsilon} = E(r)\cdot {dr
\over c} {1 \over 2\pi}  D f(\omega_{2},T_{2})d \omega_{2}.
\end{equation}
After a collision of a photon with a pair, a virtual photon will
have a velocity $v_{2}:$ $v_{2}/c=(E-2\epsilon)/(E+2\epsilon)$,
and a mass $m_{2}$: $m_{2}c^{2}=2\sqrt {2\epsilon E}$.
\par For the case of subsection 2.1, after collisions with pairs:
$\Delta E=4 \epsilon$, $\Delta \tau \geq \hbar/8\epsilon$, and we
get:
\begin{equation}
\Delta t(2\epsilon)\geq \hbar/2 \cdot {1 \over {E+ 2\epsilon} }.
\end{equation}
Then due to single gravitons and pairs:
\begin{equation}
dt_{2}(\epsilon)=dt'(\epsilon)+dt(2\epsilon)\geq
dt(\epsilon)-\hbar/2 \cdot{\epsilon E \over
{(E+\epsilon)(E+2\epsilon)}} \cdot {dr \over c} {1 \over 2\pi}D
f(\omega_{2},T_{2})d \omega_{2},
\end{equation}
where $dt'(\epsilon)$ is a reduced contribution of single
gravitons, $dt(\epsilon)$ is its full contribution corresponding
to formula (4). We see that if one takes into account graviton
pairing, the estimate of delay became smaller. So as
$${\epsilon E / {(E+\epsilon)(E+2\epsilon)}}\rightarrow 0$$ by $\epsilon/E
\rightarrow 0,$ we have for the maximal delay in this case:
$t_{2\infty}(r)\rightarrow t_{\infty}(r)$, i.e. the maximal delay
is the same as in subsection 2.1.
\par Repeating the above procedure for the case of subsection 2.2,
we shall get:
\begin{equation}
t_{2}(r)=[1+(1-1/\sqrt 2)\cdot (T_{2}/T)^{9/2}]\cdot t(r)\simeq
1.028\cdot t(r),
\end{equation}
where $t_{2}(r)$ takes into account graviton pairing, and $t(r)$
is described by formula (16). In this case, the full delay is
bigger on about $2.8 \%$ than for single gravitons.
\section[7]{Conclusion}
Because in this model the propagation time for photons as a
function of redshift is: $t(z)= r(z)/c=1/H\cdot \ln(1+z)$, the
ratio of the average time delay of photons to their propagation
time is equal approximately to $10^{-28} $ and doesn't depend on
$z$ in the first considered case. This very small quantity
characterizes the degree of Lorentz violation in the model for the
usually accepted manner of the lifetime evaluation. Even for
remote astrophysical sources time-lags will be of the order of
tens picoseconds, i.e. unmeasurable,  and one may consider Lorentz
symmetry as an exact one for any laboratory experiment. If the
second considered case is realized in the nature, one should
initially evaluate the free parameter of the model $\Delta
\tau_{0}$ from observations.
\par Some preliminary results of this work were used in my paper
\cite{501}.

\end{document}